**THE KEYNESIAN THEORY AND THE MANUFACTURED INDUSTRY IN PORTUGAL**

**Vitor João Pereira Domingues Martinho**

Unidade de I&D do Instituto Politécnico de Viseu

Av. Cor. José Maria Vale de Andrade

Campus Politécnico

3504 - 510 Viseu

**(PORTUGAL)**

**e-mail:** vdmartinho@esav.ipv.pt

# THE KEYNESIAN THEORY AND THE MANUFACTURED INDUSTRY IN PORTUGAL

**ABSTRACT**

About the economic growth the Keynesian theorists defend circular and cumulative processes, benefiting the rich localities and harming the poorest, without external interventions. In these processes the Verdoorn law has an important role. For Verdoorn (1949) the productivity growth rate is endogenous and depends of the output growth rate, capturing dynamic contexts, endogeneity of the factors and increasing economies of scale, namely in the industry. This relationship later becomes the second law of Kaldor (1966 and 1967). For Portugal there are few works or none, than those of the author, with the Verdoorn law. In this way, seem important analyze this relationship for the manufactured industry of the Portuguese regions and conclude about these contexts in Portugal. It was used data from two periods, 1986-1994 and 1995-1999, and panel data econometric methods. The two periods is to capture the effect of the Portuguese entrance in the European Economic Community and of the first Community Support Framework (1989-1993) for Portugal. As main conclusion, for the two periods, it is verified strong increasing returns in the manufactured industry and as consequence regional divergence of this sector.

## 1. INTRODUCTION

There are many works with the Verdoorn law to several countries, in different perspectives, some at regional level, another at sectoral level and other at national level. This law is used, yet in different models and with distinct econometric techniques.

For example, Dall´Erba et al. (2009) using this relationship studied the impact of the European structural funds in the manufactured industry of 145 regions of the European Union for the period 1989-2004. It was considered the possibility of spatial autocorrelation and the potential endogeneity of the independent variables. The results support the considerations of the Verdoorn relationship.

In a point of view more theoretical Llerena and Lorentz (2003) and Lorentz (2009) analyzed the micro-foundations of the Verdoorn law. They found that the dynamics scales economies emerge because the characteristics of the model and the conditions at micro level influence the economies of scale. This is an interesting conclusion, because this is in line of the new economic geography which defends, also, processes circular and cumulative, but argue increasing returns not at macro level, but at firm level.

The discussions about constant or increasing returns to scale still remain in our days, namely between the Neoclassical theorists and the Keynesian researchers, respectively. Angeriz et al. (2008) analyzed the level of increasing returns to scale in the European manufactured industry, considering the Verdoorn law, and the results are consistent with the idea of increasing scales economies. Madura (2009), considering the Kaldor frameworks, analyzed the economic growth pattern of 11 ALADI countries, with panel data, and from 1980 to 2007. The results are consistent with the predicted by Kaldor and were confirmed the ideas of increasing returns to scale. The validity of the Verddorn relationship was, also, tested by Ofria and Millemaci (2010) for some developed countries, for the period 1973-2006. The results show which the Verdoorn coefficient is stable during the period and the capital growth and the labor cost growth are not relevant to the explanation of the labor productivity growth. The evaluation of the Verdoorn law was realized by Knell (2004) in ten developed countries, during the nineties, with results coherent with the theory, namely when this law is tested on a cross-section of industries in a certain country. Basu and Foley (2011) investigated the relationship between the employment and the output from 1948 to 2010, in the United States of America, at aggregated level and some industry level, using the Verdoorn relationship and concluded about diminution of the industry importance to the service sector, with changes in the coefficients obtained.

## 2. VERDOORN MODEL

There are many specifications of the Verdoorn law, but following it will be presented the simplest model for this relationship. It is this equation that will be used in the next sections to analyze the data and to do some estimations with panel data and some recent econometric techniques.

$$p_{it} = a + bq_{it}, \text{Verdoorn model}$$

where $p_{it}$ and $q_{it}$ are the growth rates of labor productivity and of output in the industrial sector, respectively, of the economy i and in the time t.

## 3. DATA ANALYSIS

The data for the output and for the productivity were obtained from Eurostat (Eurostat Regio of Statistics), for the period 1986-1994, and were found in the INE (Statistics Portugal), for the period 1995-1999. The data are about the manufactured industry (Ferrous and non-ferrous ores and metals, other than radioactive; Non-metallic minerals and mineral products; Chemical products; Metal products, machinery, equipment and electrical goods; Transport equipment; Food, beverages, tobacco; Textiles and clothing, leather and footwear; Paper and printing products; and Products of various industries, respectively) and are disaggregated at the seven Portuguese NUTs II level.

Observing the data, in level, for the first period, represented by the figures in the annex I it is verified that in the Norte the textile industry had great importance, but the productivity there of this industry is low relatively to the others industries. Otherwise this industry had low productivity in the others NUTs II. This is because the specificity of the textile industry, many dependent yet, in the eighties and nineties, in Portugal, of the labor and, as consequence, with low increasing returns to scale. In this way we can understand the context of the NUTs II Norte. Comparatively, with the Centro and Lisboa e Vale do Tejo, the Norte had lower productivity, despite having bigger output. The Centro had mainly mineral, equipment, food and textile industries, principally the last two sectors. The food industry is the principal manufactured sector in Lisboa e Vale do Tejo, as well in the others Portuguese regions. Alentejo and Algarve had some importance of others industries, namely the metal, mineral and chemical sectors. The chemical, food and paper industries are the manufactured sectors with great productivity in the Portuguese regions. In general, Lisboa e Vale do Tejo had the biggest productivity.

From the annex II, for the second period, it is observed, more or less, the same that it is found for the first period, but of referring the fact of the metal products, machinery, equipment and electrical goods industry appear as being one of the manufactured sector with great productivity, sign of modernization.

Analyzing the figures in the annex III and IV seem that it is found more correlation between the growth rate of the productivity and the output in the first period than in the second period. But, in general it is observed some correlation among the two variables.

## 4. ESTIMATIONS RESULTS

Analyzing the results obtained with different methods and econometric techniques, presented in the following tables, they are very similar with those found by Martinho (2011) for the five Continental Portuguese regions.

From the table 1, for the first period, it is observed that the mineral and transport industries are the manufactured sectors where the Verdoorn coefficient is bigger. Sign of innovation and great increasing returns to scale. Namely the transport industry, we know which this industry, in Portugal, exists because the direct foreign investment and is a sector with large modernization. The coefficient for the paper industry is exaggeratedly high, because is bigger than one, but this is predicted by the theory, and sometimes the situations happens, sign of strong increasing returns to scale. The textile industry do not show significance statistic for the coefficient, evidence of no existence of increasing returns to scale, as expected, considering the specificities of this industry in Portugal. The situation of the equipments and electric goods industry, with coefficients not statistically significant, is a surprise, but this context occurs maybe because the same causes presented to the textile sector.

In the table 1, as well in the table 2, it is verified that the constant coefficient in the majority of the industries do not has significance statistic. From this context are possible draw two conclusions: the first is that the fixed effects have no importance, and this was proven by the tests for the existence of these effects; second the growth rate of the output is enough to explain the growth rate of the productivity, as predicted by Kaldor.

**Table 1. Results from the Verdoorn equation for the Portuguese NUTs II manufacturing, during the period 1986-1994**

|  | Const.[1] | Coef.[2] | F/Wald(mod.)[3] | F(Fe_OLS)[4] | Corr(u_i)[5] | F(Re_OLS)[6] | Hausman[7] | $R^2$ [8] | N.O.[9] | N.I.[10] |
|---|---|---|---|---|---|---|---|---|---|---|
| **Metal industry** | | | | | | | | | | |
| FE[11] | -0.059 (-1.460) | 0.675* (8.910) | 79.310* | 0.050 | -0.309 | ------- | ------- | 0.995 | 32 | ------- |
| RE[12] | -0.059 (-1.530) | 0.667* (9.980) | 99.530* | ------- | ------- | 0.000 | 0.060 | 0.995 | 32 | ------- |
| OLS | -0.059 (-1.530) | 0.667* (9.980) | 99.530* | ------- | ------- | ------- | ------- | 0.761 | 32 | ------- |
| DPD[13] | -0.085 (-1.740) | 0.694* (8.000) | 64.010* | ------- | ------- | ------- | ------- | ------- | 24 | 20 |
| **Mineral industry** | | | | | | | | | | |
| FE[11] | -0.047 (-1.710) | 0.916* (13.740) | 188.890* | 0.220 | -0.059 | ------- | ------- | 0.919 | 48 | ------- |
| RE[12] | -0.047 (-1.810) | 0.912* (14.860) | 220.880* | ------- | ------- | 0.000 | 0.030 | 0.919 | 48 | ------- |
| OLS | -0.047 (-1.810) | 0.912* (14.860) | 220.880* | ------- | ------- | ------- | ------- | 0.824 | 48 | ------- |
| DPD[13] | -0.057 (-1.650) | 0.920* (11.920) | 142.000* | ------- | ------- | ------- | ------- | ------- | 34 | 23 |
| **Chemical industry** | | | | | | | | | | |
| FE[11] | -0.013 (-0.460) | 0.502* (7.940) | 62.970* | 0.870 | 0.093 | ------- | ------- | 0.747 | 43 | ------- |
| RE[12] | -0.013 (-0.440) | 0.516* (8.560) | 73.210* | ------- | ------- | 0.870 | 0.410 | 0.747 | 43 | ------- |
| OLS | -0.012 (-0.460) | 0.515* (8.540) | 72.990* | ------- | ------- | ------- | ------- | 0.632 | 43 | ------- |
| DPD[13] | -0.030 (-1.000) | 0.471* (6.800) | 53.190* | ------- | ------- | ------- | ------- | ------- | 29 | 23 |
| **Equipments industry** | | | | | | | | | | |
| FE[11] | -0.037 (-1.420) | 0.081 (0.770) | 0.590 | 1.580 | 0.421 | ------- | ------- | 0.940 | 47 | ------- |
| RE[12] | -0.042 | 0.214* | 4.870* | ------- | ------- | 0.000 | 11.190* | 0.940 | 47 | ------- |

|  | | | | | | | | | |
|---|---|---|---|---|---|---|---|---|---|
| | (-1.560) | (2.210) | | | | | | | |
| OLS | -0.042 | 0.214* | 4.870* | ------ | ------ | ------ | ------ | 0.078 | 47 | ------ |
| | (-1.560) | (2.210) | | | | | | | |
| DPD[13] | -0.057* | -0.125 | 1.710 | ------ | ------ | ------ | ------ | ------ | 33 | 23 |
| | (-2.490) | (-1.290) | | | | | | | |
| Transport industry | | | | | | | | | |
| FE[11] | -0.038 | 0.883* | 34.710* | 0.360 | -0.243 | ------ | ------ | 0.484 | 45 | ------ |
| | (-0.770) | (5.890) | | | | | | | |
| RE[12] | -0.038 | 0.830* | 36.310* | ------ | ------ | 0.000 | 0.810 | 0.484 | 45 | ------ |
| | (-0.810) | (6.030) | | | | | | | |
| OLS | -0.038 | 0.830* | 36.310* | ------ | ------ | ------ | ------ | 0.445 | 45 | ------ |
| | (-0.810) | (6.030) | | | | | | | |
| DPD[13] | -0.054 | 0.779* | 27.400* | ------ | ------ | ------ | ------ | ------ | 32 | 23 |
| | (-0.920) | (3.990) | | | | | | | |
| Food industry | | | | | | | | | |
| FE[11] | 0.004 | 0.723* | 22.100* | 0.210 | -0.051 | ------ | ------ | 0.576 | 48 | ------ |
| | (0.210) | (4.700) | | | | | | | |
| RE[12] | 0.005 | 0.714* | 27.060* | ------ | ------ | 0.000 | 0.010 | 0.576 | 48 | ------ |
| | (0.250) | (5.200) | | | | | | | |
| OLS | 0.005 | 0.714* | 27.060* | ------ | ------ | ------ | ------ | 0.357 | 48 | ------ |
| | (0.250) | (5.200) | | | | | | | |
| DPD[13] | 0.004 | 0.677* | 12.330* | ------ | ------ | ------ | ------ | ------ | 34 | 23 |
| | (0.150) | (3.510) | | | | | | | |
| Textile industry | | | | | | | | | |
| FE[11] | -0.006 | 0.179 | 1.980 | 0.320 | 0.194 | ------ | ------ | 0.680 | 44 | ------ |
| | (-0.150) | (1.410) | | | | | | | |
| RE[12] | -0.005 | 0.213 | 3.370 | ------ | ------ | 0.000 | 0.410 | 0.680 | 44 | ------ |
| | (-0.150) | (1.830) | | | | | | | |
| OLS | -0.005 | 0.213 | 3.370 | ------ | ------ | ------ | ------ | 0.052 | 44 | ------ |
| | (-0.150) | (1.830) | | | | | | | |
| DPD[13] | -0.020 | 0.170 | 1.100 | ------ | ------ | ------ | ------ | ------ | 30 | 23 |
| | (-0.440) | (1.040) | | | | | | | |
| Paper industry | | | | | | | | | |
| FE[11] | -0.054* | 1.130* | 169.270* | 0.330 | -0.018 | ------ | ------ | 0.809 | 48 | ------ |
| | (-2.070) | (13.010) | | | | | | | |
| RE[12] | -0.053* | 1.128* | 188.210* | ------ | ------ | 0.000 | 0.010 | 0.809 | 48 | ------ |
| | (-2.160) | (13.720) | | | | | | | |
| OLS | -0.053* | 1.128* | 188.210* | ------ | ------ | ------ | ------ | 0.799 | 48 | ------ |
| | (-2.160) | (13.720) | | | | | | | |
| DPD[13] | -0.059 | 1.140* | 129.400* | ------ | ------ | ------ | ------ | ------ | 34 | 23 |
| | (-1.890) | (11.340) | | | | | | | |
| Several industry | | | | | | | | | |
| FE[11] | -0.014 | 0.561* | 71.560* | 0.110 | 0.084 | ------ | ------ | 0.943 | 48 | ------ |
| | (-0.580) | (8.460) | | | | | | | |
| RE[12] | -0.014 | 0.566* | 83.140* | ------ | ------ | 0.000 | 0.040 | 0.943 | 48 | ------ |
| | (-0.620) | (9.120) | | | | | | | |
| OLS | -0.014 | 0.566* | 83.140* | ------ | ------ | ------ | ------ | 0.636 | 48 | ------ |
| | (-0.620) | (9.120) | | | | | | | |
| DPD[13] | -0.028 | 0.504* | 56.550* | ------ | ------ | ------ | ------ | ------ | 34 | 23 |
| | (-0.980) | (5.980) | | | | | | | |

Note: 1, Constant; 2, Coefficient; 3, Test F for fixed effects model and test Wald for random effects and dynamic panel data models; 4, Test F for fixed effects or OLS (Ho is OLS); 5, Correlation between errors and regressors in fixed effects; 6, Test F for random effects or OLS (Ho is OLS); 7, Hausman test (Ho is GLS); 8, R square; 9, Number of observations; 10, Number of instruments;, 11, Fixed effects model; 12, Random effects model; 13, Dynamic panel data model; *, Statically significant at 5%.

The table 2 shows that the industries with more increasing returns to scale, in the second period, are the chemical and paper sectors. The textile industry improves lightly and the equipments industry remains in the same context like that presented in the first period.

Table 2. Results from the Verdoorn equation for the Portuguese NUTs II manufacturing, during the period 1995-1999

| | Const.[1] | Coef.[2] | F/Wald(mod.)[3] | F(Fe_OLS)[4] | Corr(u_i)[5] | F(Re_OLS)[6] | Hausman[7] | $R^2$[8] | N.O.[9] | N.I.[10] |
|---|---|---|---|---|---|---|---|---|---|---|
| Metal industry | | | | | | | | | | |
| FE[11] | -0.058 | 0.800* | 13.910* | 1.470 | -0.446 | ------ | ------ | 0.410 | 28 | ------ |
| | (-1.870) | (3.730) | | | | | | | | |
| RE[12] | -0.034 | 0.606* | 13.400* | ------ | ------ | 0.020 | 2.030 | 0.410 | 28 | ------ |
| | (-1.220) | (3.660) | | | | | | | | |
| OLS | -0.030 | 0.574* | 12.780* | ------ | ------ | ------ | ------ | 0.304 | 28 | ------ |
| | (-1.160) | (3.580) | | | | | | | | |
| DPD[13] | -0.066 | 0.847* | 13.470* | ------ | ------ | ------ | ------ | ------ | 14 | 5 |
| | (-1.760) | (3.610) | | | | | | | | |
| **Mineral industry** | | | | | | | | | | |
| FE[11] | 0.009 (0.690) | 0.756* (6.270) | 39.290* | 0.520 | 0.111 | ------- | ------- | 0.702 | 28 | ------- |
| RE[12] | 0.007 (0.640) | 0.779* (6.990) | 48.890* | ------- | ------- | 0.000 | 0.260 | 0.702 | 28 | ------- |
| OLS | 0.007 (0.640) | 0.779* (6.990) | 48.890* | ------- | ------- | ------- | ------- | 0.640 | 28 | ------- |
| DPD[13] | 0.014 (0.540) | 0.787* (3.300) | 28.320* | ------- | ------- | ------- | ------- | ------- | 14 | 5 |
| **Chemical industry** | | | | | | | | | | |
| FE[11] | -0.010 (-0.360) | 0.869* (5.410) | 29.280* | 0.450 | -0.027 | ------- | ------- | 0.614 | 28 | ------- |
| RE[12] | -0.010 (-0.390) | 0.862* (6.200) | 38.420* | ------- | ------- | 0.000 | 0.010 | 0.614 | 28 | ------- |
| OLS | -0.010 (-0.390) | 0.862* (6.200) | 38.420* | ------- | ------- | ------- | ------- | 0.581 | 28 | ------- |
| DPD[13] | -0.029 (-0.700) | 1.050* (4.160) | 18.550* | ------- | ------- | ------- | ------- | ------- | 14 | 5 |
| **Equipments industry** | | | | | | | | | | |
| FE[11] | 0.014 (0.700) | -0.023 (-0.140) | 0.020 | 0.410 | -0.082 | ------- | ------- | 0.014 | 28 | ------- |
| RE[12] | 0.013 (0.720) | -0.004 (-0.030) | 0.000 | ------- | ------- | 0.000 | 0.040 | 0.014 | 28 | ------- |
| OLS | 0.013 (0.720) | -0.004 (-0.030) | 0.000 | ------- | ------- | ------- | ------- | 0.038 | 28 | ------- |
| DPD[13] | 0.064 (0.710) | -0.051 (-0.180) | 0.680 | ------- | ------- | ------- | ------- | ------- | 14 | 5 |
| **Transport industry** | | | | | | | | | | |
| FE[11] | -0.006 (-0.170) | 0.630* (3.940) | 15.550* | 0.070 | -0.089 | ------- | ------- | 0.676 | 28 | ------- |
| RE[12] | -0.004 (-0.150) | 0.621* (4.560) | 20.780* | ------- | ------- | 0.000 | 0.010 | 0.676 | 28 | ------- |
| OLS | -0.004 (-0.150) | 0.621* (4.560) | 20.780* | ------- | ------- | ------- | ------- | 0.423 | 28 | ------- |
| DPD[13] | -0.031 (-0.490) | 0.209 (0.540) | 1.080 | ------- | ------- | ------- | ------- | ------- | 14 | 5 |
| **Food industry** | | | | | | | | | | |
| FE[11] | 0.040 (1.730) | 0.593 (1.910) | 3.630 | 0.330 | 0.200 | ------- | ------- | 0.672 | 28 | ------- |
| RE[12] | 0.036 (1.700) | 0.679* (2.490) | 6.200* | ------- | ------- | 0.000 | 0.330 | 0.672 | 28 | ------- |
| OLS | 0.036 (1.700) | 0.679* (2.490) | 6.200* | ------- | ------- | ------- | ------- | 0.161 | 28 | ------- |
| DPD[13] | 0.110* (3.100) | 0.413 (1.260) | 12.870* | ------- | ------- | ------- | ------- | ------- | 14 | 5 |
| **Textile industry** | | | | | | | | | | |
| FE[11] | 0.024 (1.720) | 0.438* (2.690) | 7.220* | 0.350 | -0.431 | ------- | ------- | 0.265 | 28 | ------- |
| RE[12] | 0.024 (1.910) | 0.345* (2.770) | 7.690* | ------- | ------- | 0.000 | 0.780 | 0.265 | 28 | ------- |
| OLS | 0.024 (1.910) | 0.345* (2.770) | 7.690* | ------- | ------- | ------- | ------- | 0.199 | 28 | ------- |
| DPD[13] | -0.016 (-0.480) | 1.240* (2.290) | 7.030* | ------- | ------- | ------- | ------- | ------- | 14 | 5 |
| **Paper industry** | | | | | | | | | | |
| FE[11] | -0.021 (-0.970) | 0.869* (4.280) | 18.350* | 0.280 | 0.072 | ------- | ------- | 0.817 | 28 | ------- |
| RE[12] | -0.021 (-1.060) | 0.886* (5.560) | 30.920* | ------- | ------- | 0.000 | 0.020 | 0.817 | 28 | ------- |
| OLS | -0.021 (-1.060) | 0.886* (5.560) | 30.920* | ------- | ------- | ------- | ------- | 0.526 | 28 | ------- |
| DPD[13] | 0.004 (0.170) | 0.906* (3.700) | 16.080* | ------- | ------- | ------- | ------- | ------- | 14 | 5 |
| **Several industry** | | | | | | | | | | |
| FE[11] | 0.024 (0.550) | 1.129* (5.350) | 28.640* | 1.140 | -0.092 | ------- | ------- | 0.663 | 28 | ------- |
| RE[12] | 0.027 (0.580) | 1.096* (6.310) | 39.810* | ------- | ------- | 0.000 | 0.080 | 0.663 | 28 | ------- |
| OLS | 0.028 (0.660) | 1.089* (6.360) | 40.500* | ------- | ------- | ------- | ------- | 0.594 | 28 | ------- |
| DPD[13] | 0.088 (0.960) | 1.196* (3.970) | 15.910* | ------- | ------- | ------- | ------- | ------- | 14 | 5 |

**Note: 1, Constant; 2, Coefficient; 3, Test F for fixed effects model and test Wald for random effects and dynamic panel data models; 4, Test F for fixed effects or OLS (Ho is OLS); 5, Correlation between errors and regressors in fixed effects; 6, Test F for random effects or OLS (Ho is OLS); 7, Hausman test (Ho is GLS); 8, R square; 9, Number of observations; 10, Number of instruments;, 11, Fixed effects model; 12, Random effects model; 13, Dynamic panel data model; *, Statically significant at 5%.**

The table 3 presents the results, of several tests, for the stationary analysis of the data about the growth rate of the output and of the productivity, in the first period. From the values of the tests it is possible to conclude about the no volatility of the data. However in some cases, like the metal and mineral industries situation, it was needed to use more than one lag to obtain tests with statistical significance. On other hand, seems that the output presents more stationary data.

**Table 3. Tests of the stationary with the Fisher-type unit root-test based on Philips-Perron tests, for the variables used in the Verdoorn equation (NUTs II, 1986-1994)**

| Metal industry | | |
|---|---|---|
| | $p^b$ | $q^b$ |
| Inverse[1] | 16.107* | 15.209* |
| Inverse[2] | -1.812 | -1.833* |
| Inverse[3] | -1.854* | -1.825* |
| **Mineral industry** | | |
| | $p^b$ | $q^a$ |
| Inverse[1] | 25.299* | 104.341* |
| Inverse[2] | -1.600* | -5.399* |
| Inverse[3] | -1.679* | -10.285* |
| **Chemical industry** | | |
| | $p^a$ | $q^a$ |
| Inverse[1] | 49.759* | 41.455* |
| Inverse[2] | -4.695* | -3.604* |
| Inverse[3] | -5.505* | -4.016* |
| **Equipments industry** | | |
| | $p^a$ | $q^a$ |
| Inverse[1] | 108.230* | 42.339* |
| Inverse[2] | -6.610* | -2.888* |
| Inverse[3] | 12.151* | -3.407* |
| **Transport industry** | | |
| | $p^a$ | $q^a$ |
| Inverse[1] | 87.054* | 63.283* |
| Inverse[2] | -6.645* | -4.948* |
| Inverse[3] | -9.835* | -6.082* |
| **Food industry** | | |
| | $p^a$ | $q^a$ |
| Inverse[1] | 69.109* | 68.191* |
| Inverse[2] | -4.417* | -5.487* |
| Inverse[3] | -5.765* | -7.052* |
| **Textile industry** | | |
| | $p^a$ | $q^a$ |
| Inverse[1] | 29.383* | 73.305* |
| Inverse[2] | -2.136* | -5.884* |
| Inverse[3] | -2.086* | -7.527* |
| **Paper industry** | | |
| | $p^a$ | $q^a$ |
| Inverse[1] | 31.808* | 37.811* |
| Inverse[2] | -2.585* | -2.389* |
| Inverse[3] | -2.725* | -3.031* |
| **Several industry** | | |
| | $p^a$ | $q^a$ |
| Inverse[1] | 54.615* | 132.030* |
| Inverse[2] | -3.452* | -7.167* |
| Inverse[3] | 4.997* | -13.692* |

[1] Inverse chi-squared (P); [2] Inverse normal (Z); [3] Inverse logit t (L*); [a] ADF regression with one lag; [b] ADF regression with more than one lag; * Statistically significant at 5%.

The conclusions about the volatility of the data for the second period, table 4, are more or less similar with those presented before for the first period.

**Table 4. Tests of the stationary with the Fisher-type unit root-test based on Philips-Perron tests, for the variables used in the Verdoorn equation (NUTs II, 1995-1999)**

| Metal industry | | |
|---|---|---|
| | $p^a$ | $q^b$ |
| Inverse[1] | 48.662* | 53.234* |
| Inverse[2] | -2.694* | -1.792* |
| Inverse[3] | -4.370* | -4.182* |
| **Mineral industry** | | |
| | $p^a$ | $q^a$ |
| Inverse[1] | 81.513* | 96.708* |
| Inverse[2] | -5.701* | -4.320* |
| Inverse[3] | -8.135* | -8.865* |
| **Chemical industry** | | |
| | $p^a$ | $q^a$ |
| Inverse[1] | 81.981* | 22.923* |
| Inverse[2] | -3.068* | -1.829* |
| Inverse[3] | -7.621* | -1.836* |
| **Equipments industry** | | |
| | $p^a$ | $q^a$ |
| Inverse[1] | 30.303* | 29.504* |
| Inverse[2] | -1.392 | -2.005* |
| Inverse[3] | -1.696* | -2.465* |
| **Transport industry** | | |
| | $p^a$ | $q^a$ |
| Inverse[1] | 183.842* | 83.427* |
| Inverse[2] | -8.772* | -4.282* |
| Inverse[3] | -18.117* | -7.844* |
| **Food industry** | | |
| | $p^a$ | $q^b$ |
| Inverse[1] | 61.216* | 45.540* |
| Inverse[2] | -3.818* | -1.722* |
| Inverse[3] | -5.330* | -2.654* |
| **Textile industry** | | |
| | $p^b$ | $q^a$ |
| Inverse[1] | 33.047* | 54.060* |
| Inverse[2] | -1.689* | -3.006* |
| Inverse[3] | -2.301* | -4.639* |
| **Paper industry** | | |
| | $p^a$ | $q^a$ |
| Inverse[1] | 104.288* | 119.451* |
| Inverse[2] | -5.574* | -6.211* |
| Inverse[3] | -10.364* | -11.878* |
| **Several industry** | | |
| | $p^a$ | $q^a$ |
| Inverse[1] | 34.839* | 75.426* |
| Inverse[2] | -1.569* | -1.268 |
| Inverse[3] | -2.213* | -6.065* |

[1] Inverse chi-squared (P); [2] Inverse normal (Z); [3] Inverse logit t (L*); [a] ADF regression with one lag; [b] ADF regression with more than one lag; * Statistically significant at 5%.

## 5. CONCLUSIONS

Based on the Verdoorn law, from the data analysis presented before and from the results found with different econometric techniques and methods, presented in the tables 1 and 2, it is concluded about the importance of the food industry in the Portuguese regions. The Norte NUTs II present great specificity in the textile industry and the Centro do not presents apparent specificity. In the others NUTs II the food industry is the principal industry.

The manufacturing sector, in the Portuguese regions, from 1986 to 1999, presents significant increasing returns to scale. To stress the fact of the textile industry presents improvements, in terms of scales economies, from the first to the second period, sign of modernization, and in Portugal we know that this really happens.

The <u>metal products, machinery, equipment and electrical goods</u> industry was a surprise the fact of did not present Verdoorn coefficients with statistical significance, but will be a interesting future research to try indentify the causes of these results.

As final conclusion of referring which there were changes from the first to the second period in the values of the coefficient. This situation will be because the change in the method of collection the data, by the official statistics institutions, or because there were changes in the structures of the industries, what in some cases will be good, but in others cases maybe not. Will be important clarify this context in future researches.

In general, it is found signs of strong increasing returns to scale, for the manufacturing sector, what is good for the Portuguese economy, but if we think that this sector is principally in the littoral and that Portugal has a asymmetry between the interior and the littoral, maybe is needed to identify adjusted regional policies to improve this processes.

Annex 1

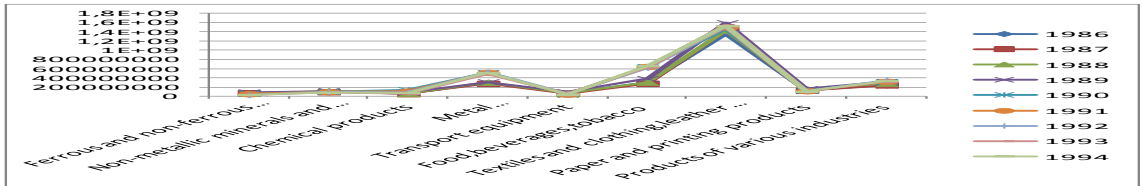
**Figure1: Manufacturing output, in euros, for Portuguese NUT II Norte (1986-1994)**

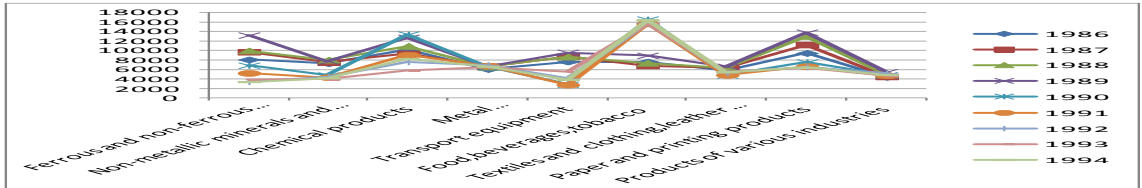
**Figure2: Manufacturing productivity, in euros, for Portuguese NUT II Norte (1986-1994)**

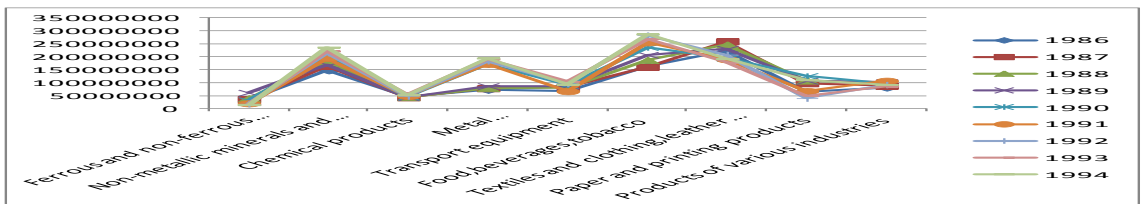
**Figure3: Manufacturing output, in euros, for Portuguese NUT II Centro (1986-1994)**

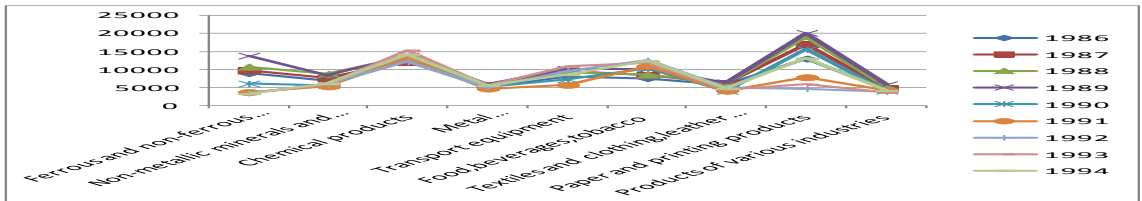
**Figure4: Manufacturing productivity, in euros, for Portuguese NUT II Centro (1986-1994)**

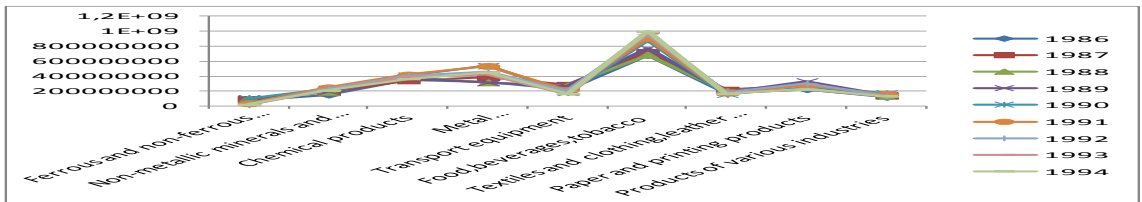
**Figure5: Manufacturing output, in euros, for Portuguese NUT II Lisboa e Vale do Tejo (1986-1994)**

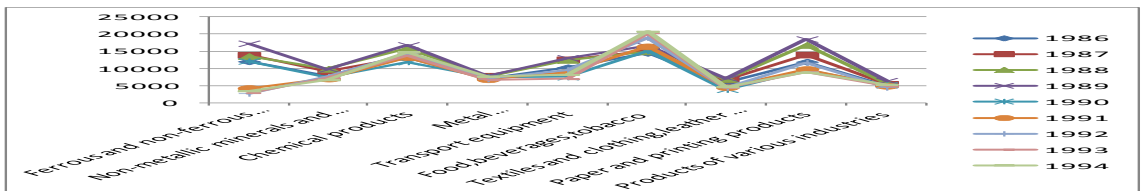
**Figure6: Manufacturing productivity, in euros, for Portuguese NUT II Lisboa e Vale do Tejo (1986-1994)**

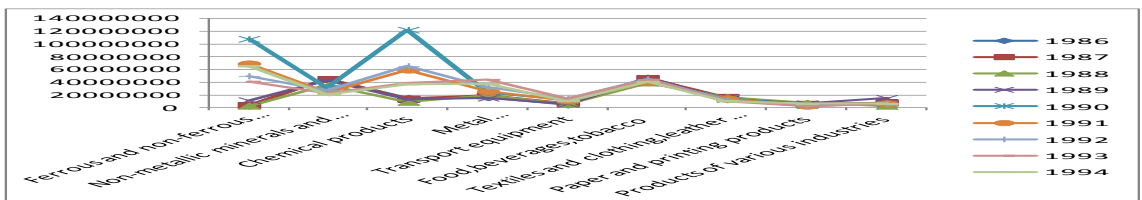
**Figure7: Manufacturing output, in euros, for Portuguese NUT II Alentejo (1986-1994)**

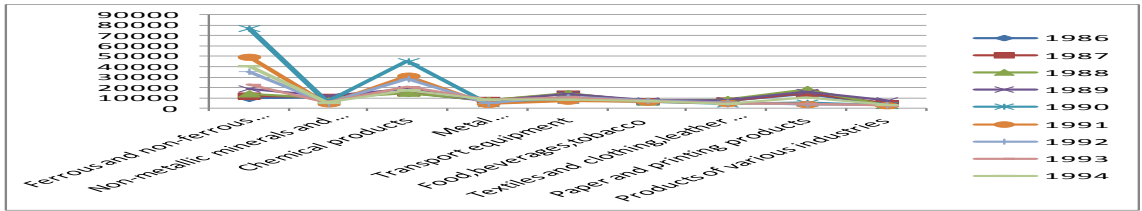
**Figure8:** Manufacturing productivity, in euros, for Portuguese NUT II Alentejo (1986-1994)

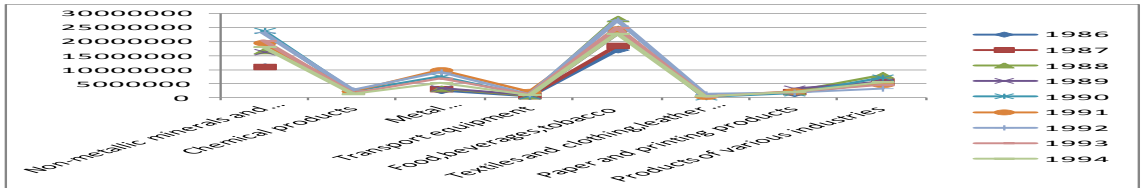
**Figure9:** Manufacturing output, in euros, for Portuguese NUT II Algarve (1986-1994)

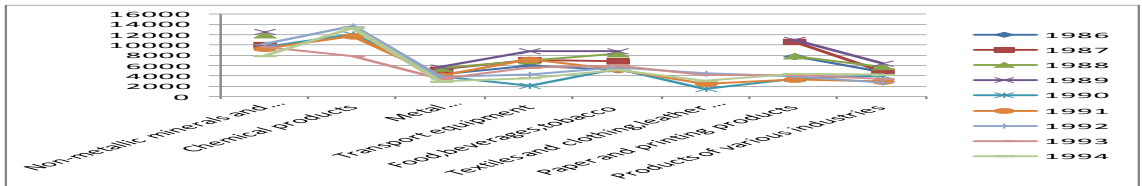
**Figure10:** Manufacturing productivity, in euros, for Portuguese NUT II Algarve (1986-1994)

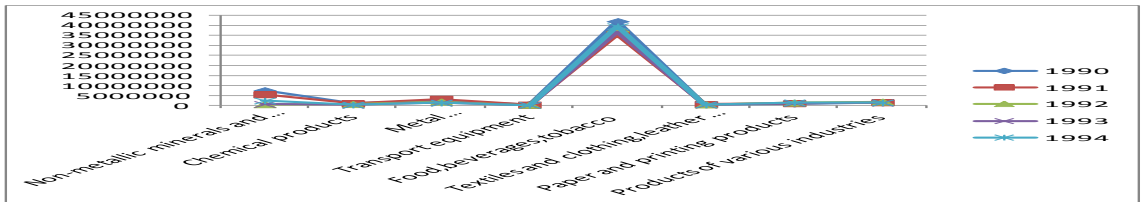
**Figure11:** Manufacturing output, in euros, for Portuguese NUT II Açores (1986-1994)

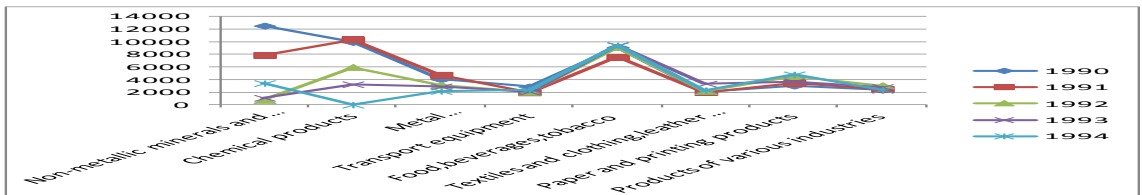
**Figure12:** Manufacturing productivity, in euros, for Portuguese NUT II Açores (1986-1994)

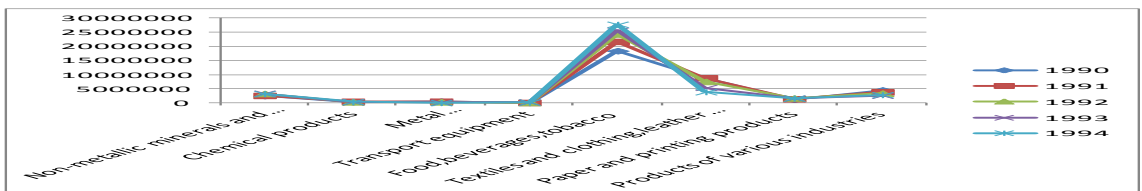
**Figure13:** Manufacturing output, in euros, for Portuguese NUT II Madeira (1986-1994)

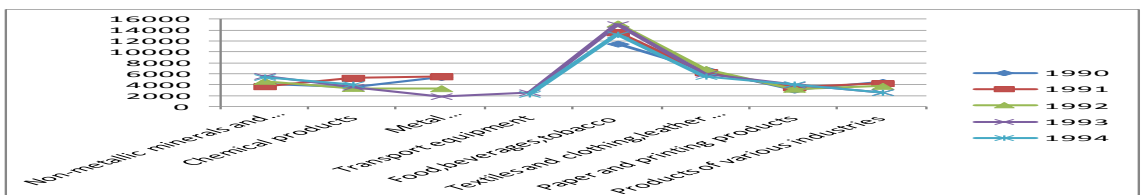
**Figure14:** Manufacturing productivity, in euros, for Portuguese NUT II Madeira (1986-1994)

Annex II

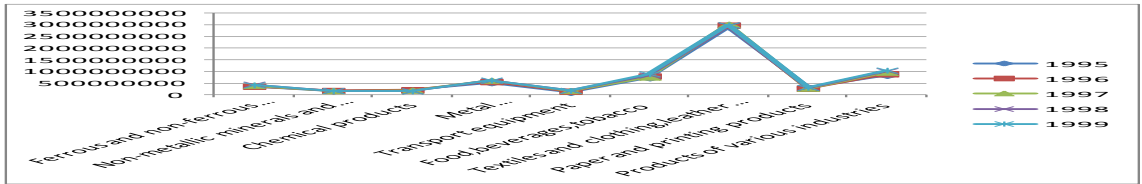

**Figure15: Manufacturing output, in euros, for Portuguese NUT II Norte (1995-1999)**

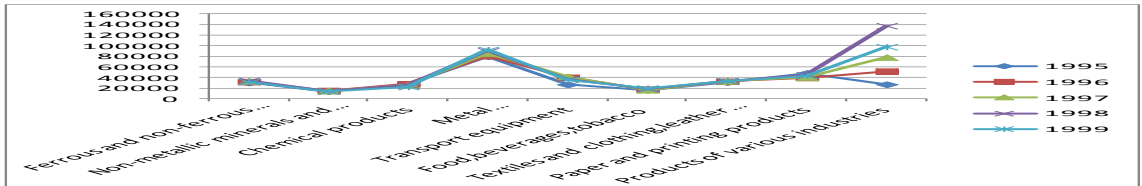

**Figure16: Manufacturing productivity, in euros, for Portuguese NUT II Norte (1995-1999)**

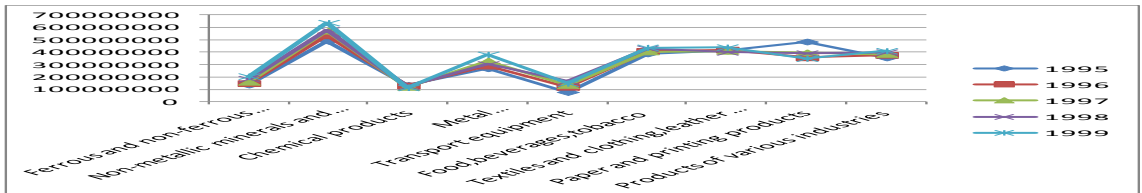

**Figure17: Manufacturing output, in euros, for Portuguese NUT II Centro (1995-1999)**

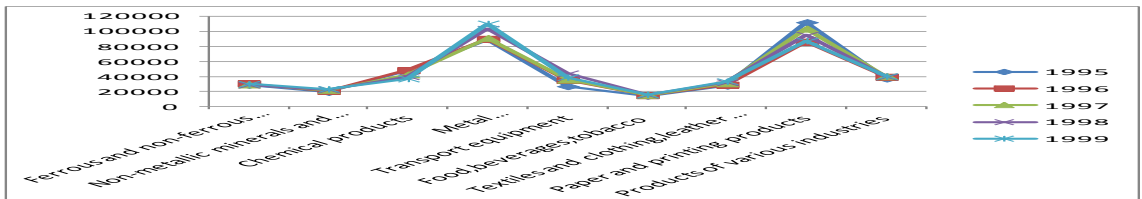

**Figure18: Manufacturing productivity, in euros, for Portuguese NUT II Centro (1995-1999)**

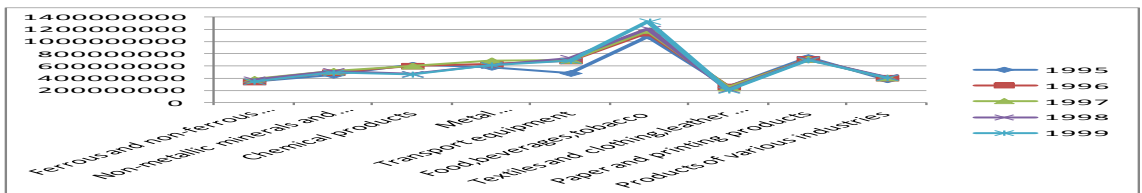

**Figure19: Manufacturing output, in euros, for Portuguese NUT II Lisboa e Vale do Tejo (1995-1999)**

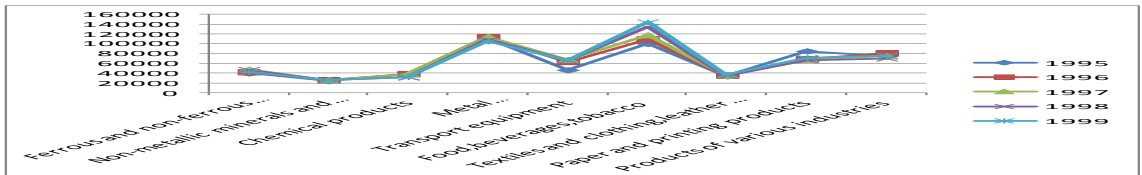

**Figure20: Manufacturing productivity, in euros, for Portuguese NUT II Lisboa e Vale do Tejo (1995-1999)**

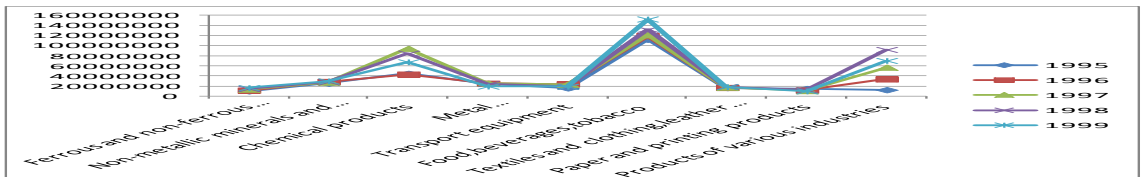

**Figure21: Manufacturing output, in euros, for Portuguese NUT II Alentejo (1995-1999)**

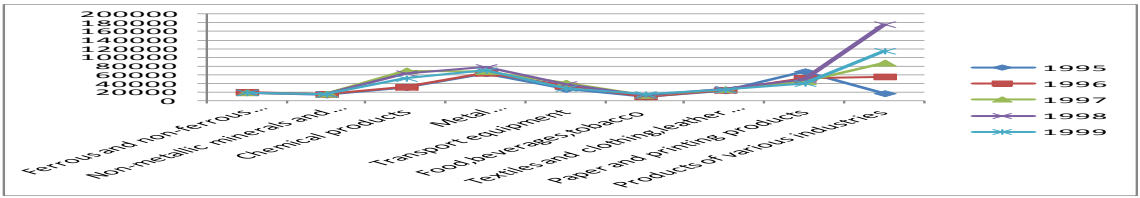
**Figure22: Manufacturing productivity, in euros, for Portuguese NUT II Alentejo (1995-1999)**

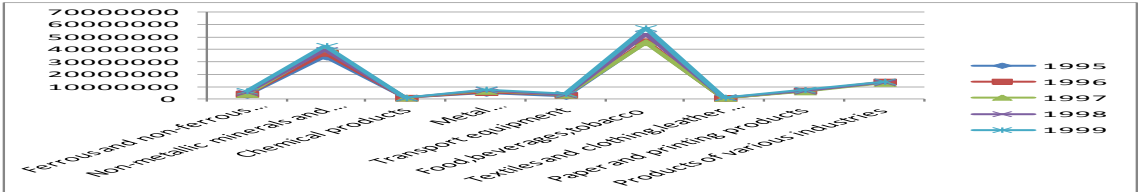
**Figure23: Manufacturing output, in euros, for Portuguese NUT II Algarve (1995-1999)**

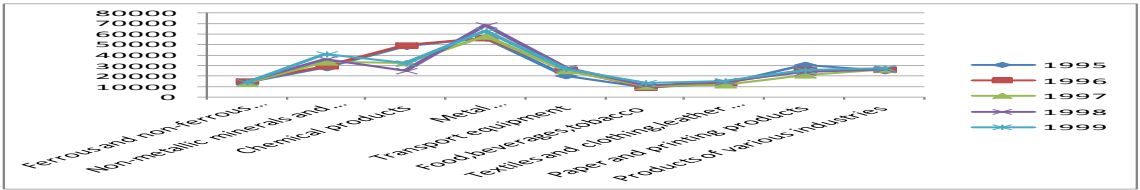
**Figure24: Manufacturing productivity, in euros, for Portuguese NUT II Algarve (1995-1999)**

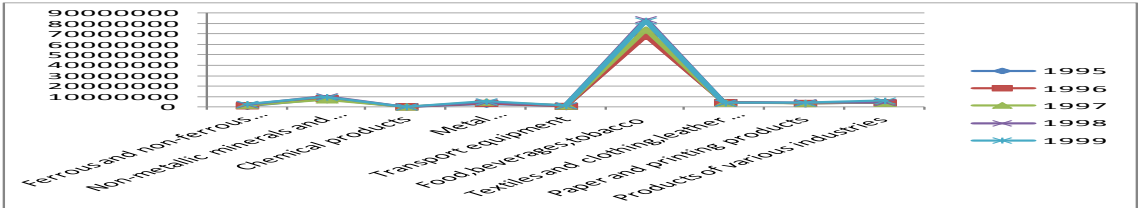
**Figure25: Manufacturing output, in euros, for Portuguese NUT II Açores (1995-1999)**

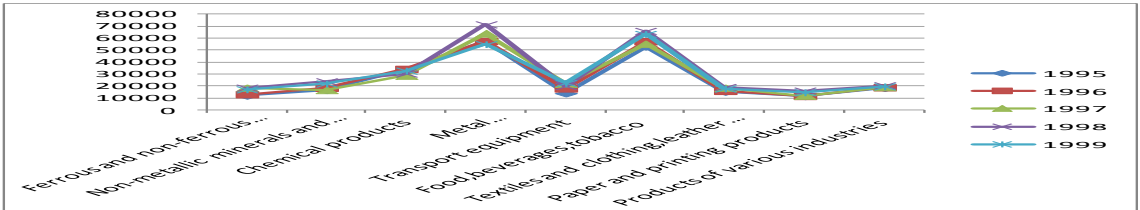
**Figure 26: Manufacturing productivity, in euros, for Portuguese NUT II Açores (1995-1999)**

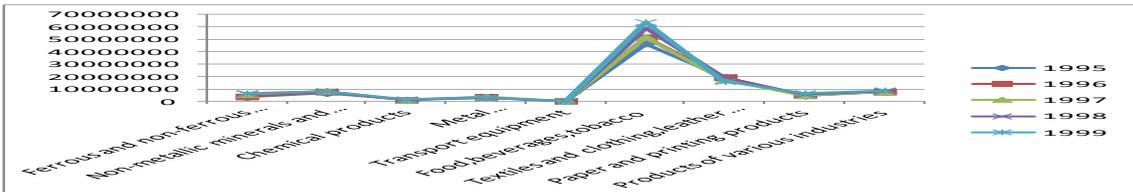
**Figure27: Manufacturing output, in euros, for Portuguese NUT II Madeira (1995-1999)**

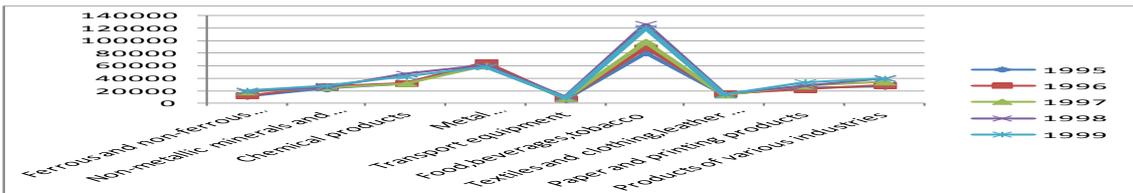
**Figure 28: Manufacturing productivity, in euros, for Portuguese NUT II Madeira (1995-1999)**

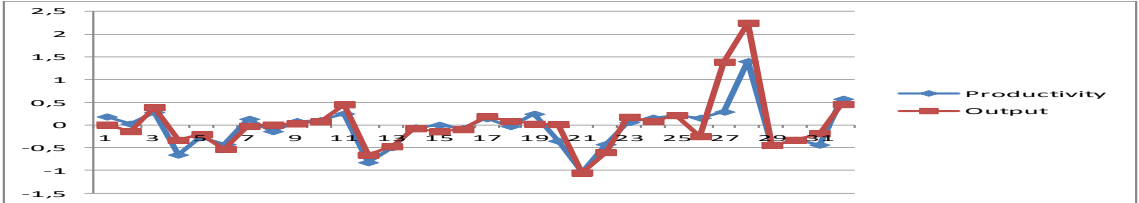
**Figure 29: Relationship between productivity and output growth rates, for the metal industry (1986-1994)**

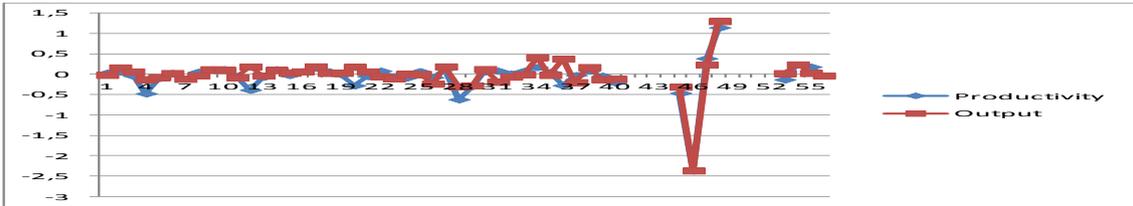
**Figure 30: Relationship between productivity and output growth rates, for the mineral industry (1986-1994)**

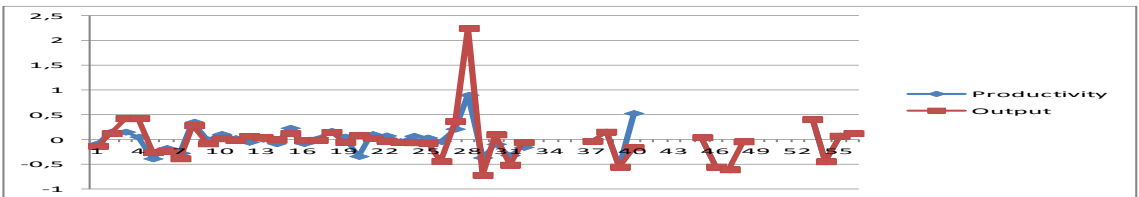
**Figure 31: Relationship between productivity and output growth rates, for the chemical industry (1986-1994)**

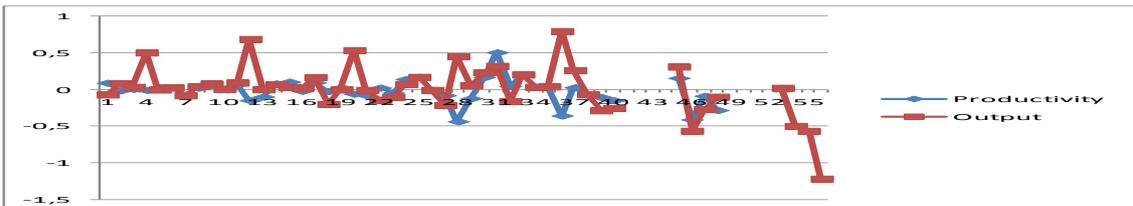
**Figure 32: Relationship between productivity and output growth rates, for the equipments industry (1986-1994)**

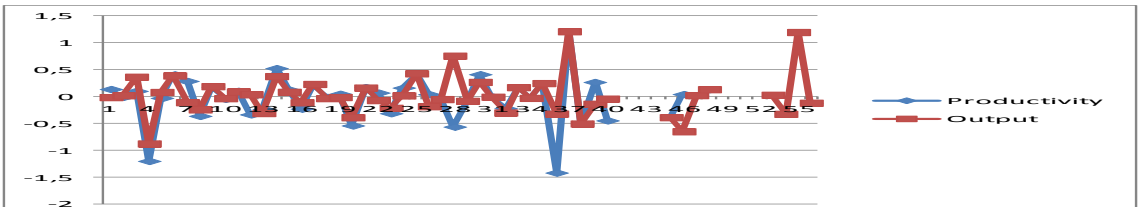
**Figure 33: Relationship between productivity and output growth rates, for the transport industry (1986-1994)**

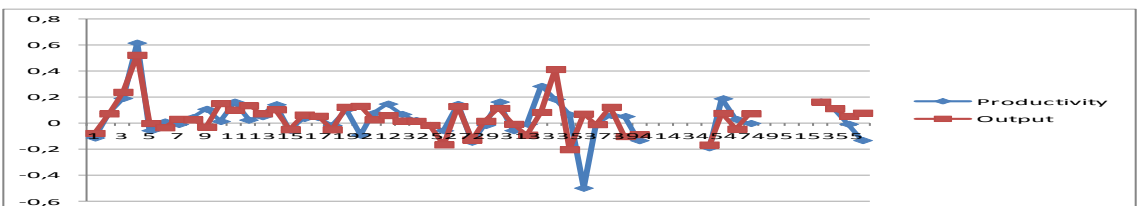
**Figure 34: Relationship between productivity and output growth rates, for the food industry (1986-1994)**

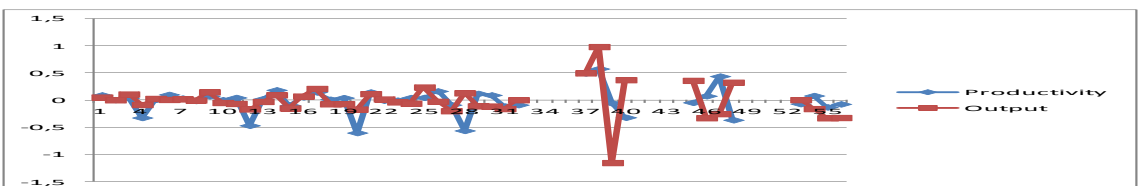
**Figure 35: Relationship between productivity and output growth rates, for the textile industry (1986-1994)**

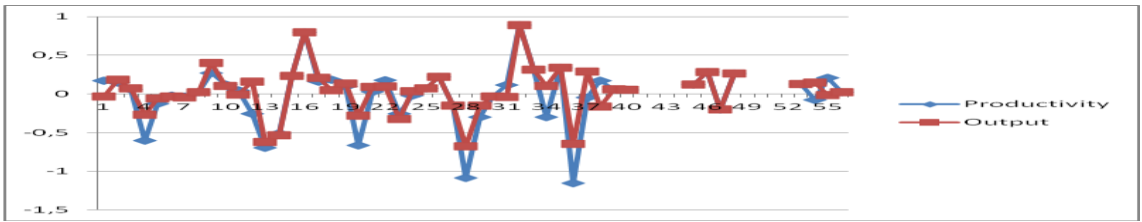
**Figure 36:** Relationship between productivity and output growth rates, for the paper industry (1986-1994)

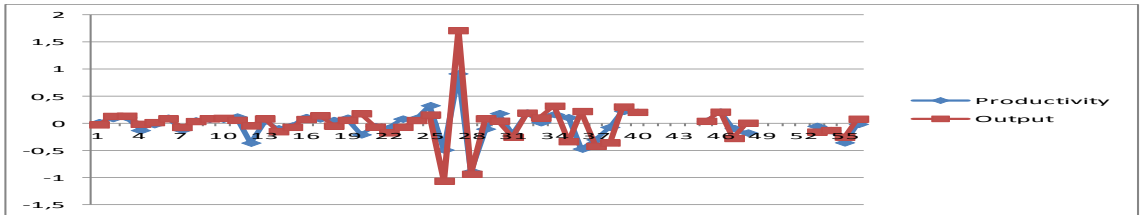
**Figure 37:** Relationship between productivity and output growth rates, for several industry (1986-1994)

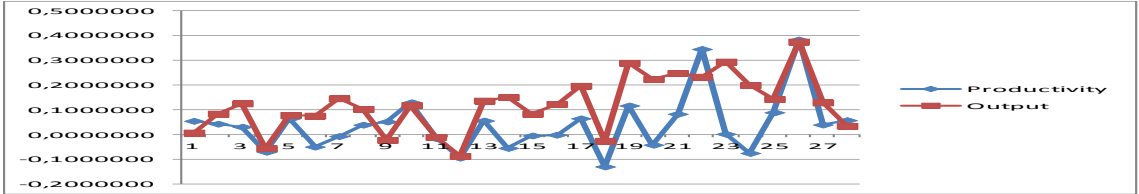
**Figure 38: Relationship between productivity and output growth rates, for metal industry (1995-1999)**

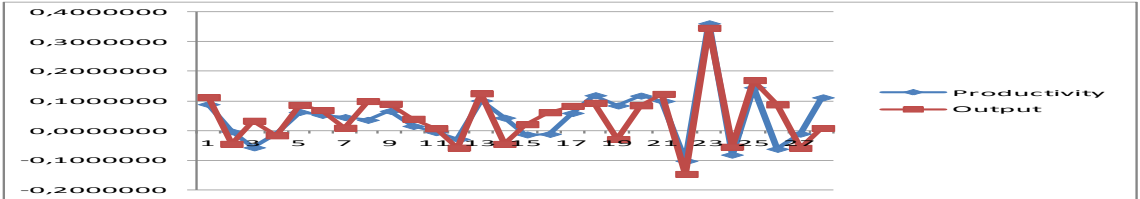
**Figure 39: Relationship between productivity and output growth rates, for mineral industry (1995-1999)**

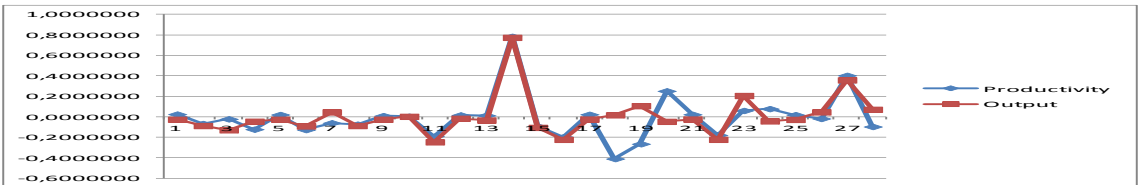
**Figure 40: Relationship between productivity and output growth rates, for chemical industry (1995-1999)**

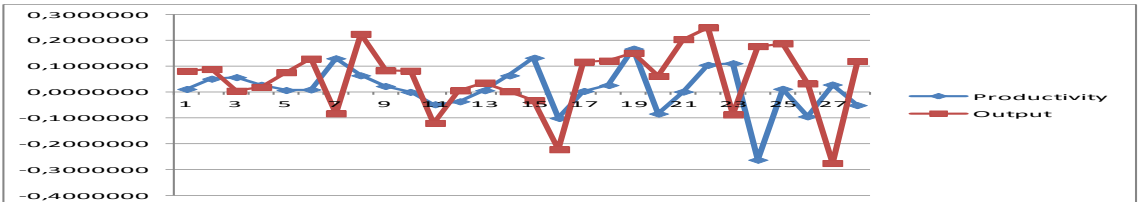
**Figure 41: Relationship between productivity and output growth rates, for equipments industry (1995-1999)**

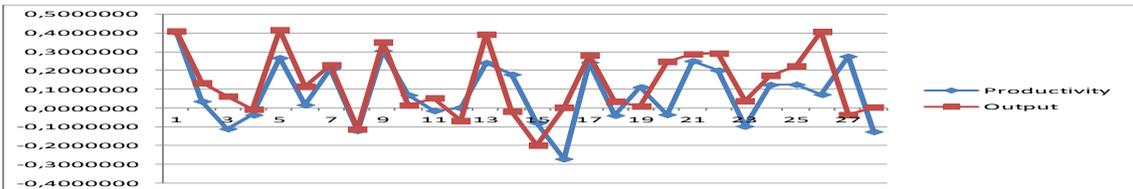
**Figure 42: Relationship between productivity and output growth rates, for transport industry (1995-1999)**

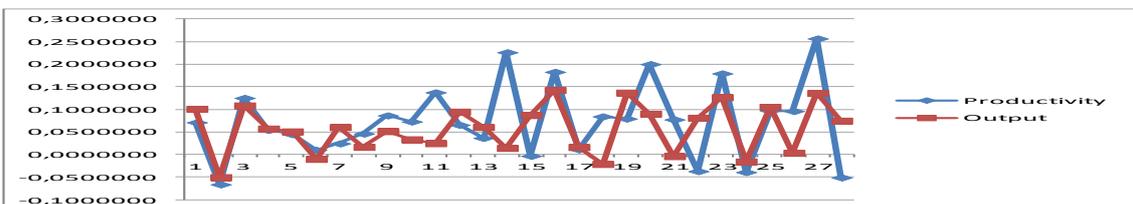
**Figure 43: Relationship between productivity and output growth rates, for food industry (1995-1999)**

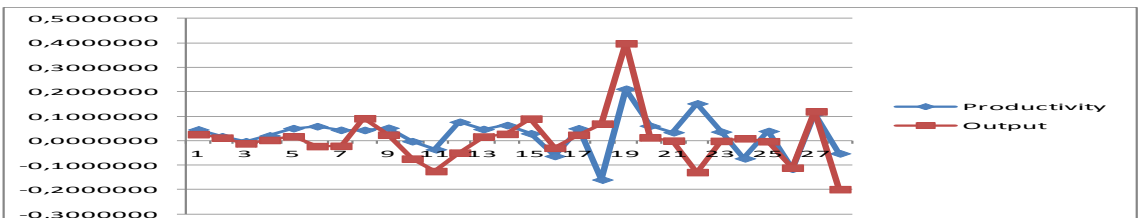
**Figure 44: Relationship between productivity and output growth rates, for textile industry (1995-1999)**

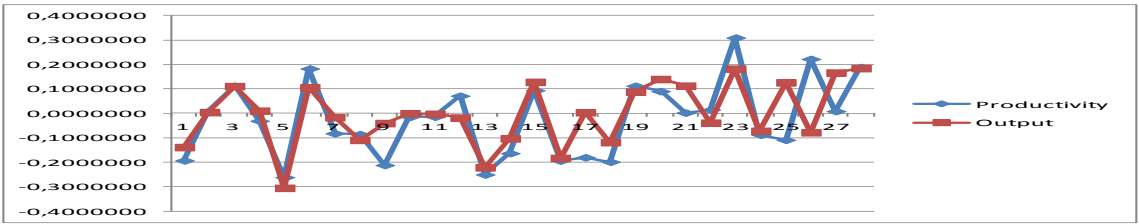

**Figure 45: Relationship between productivity and output growth rates, for paper industry (1995-1999)**

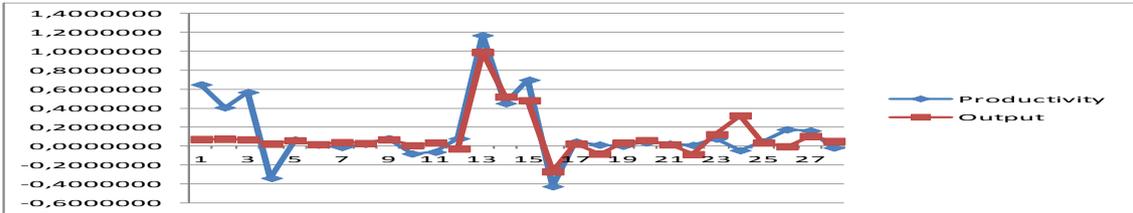

**Figure 46: Relationship between productivity and output growth rates, for several industry (1995-1999)**